\documentclass[aip, graphicx, apl, numerical,
reprint,
 amsmath,amssymb,
]{revtex4-1}

\usepackage{graphics}
\usepackage{dcolumn}
\usepackage{bm}

\begin{document}


\title{Combined photo- and electroreflectance of multijunction solar cells enabled by subcell electric coupling }

\author{D. \surname{Fuertes Marr\'on}}
\email{dfuertes@ies.upm.es}
\author{E. Barrig\'on}%
 \altaffiliation[Presently at ]{Solid State Physics and Nano-Lund, Lund University, P.\ O.\ Box 118, SE-221 00 Lund, Sweden.}
\author{M. Ochoa}
\altaffiliation[Presently at ]{Empa-Swiss Federal Laboratories for Mater. Sci. and Tech., Ueberlandstrasse 129, 8600 D\"ubendorf, Switzerland.}
\author{I. Artacho}
\affiliation{%
 Instituto de Energ\'ia Solar, Universidad Polit\'ecnica de Madrid, ETSI Telecomunicaci\'on, Ciudad Universitaria s/n, 28040 Madrid, Spain
}%

\date{\today}

\begin{abstract}
Electric coupling between subcells of a monolithically grown multijunction solar cell in short circuit allows their simultaneous and independent characterization by means of photo- and electroreflectance.\ The photovoltage generated by selective absorption of the pump beam in a given subcell during photoreflectance measurements results in reverse biasing the complementary subunits at the modulation frequency set on the pump illumination.\ Such voltage bias modulation acts then as external perturbation on the complementary subcells.\ The spectral separation of the different subcell absorption ranges permits the probe beam to record in a single spectrum the response of the complete device as a combination of photo- and electroreflectance, thereby providing access for diagnosis of subcells on an individual basis.\ This form of modulation spectroscopy is demonstrated on a GaInP/GaAs tandem solar cell.
\end{abstract}

\maketitle




Photoreflectance (PR) belongs to the family of modulation spectroscopies \cite{Card69, Poll01}, in which tiny changes in the dielectric constant of the material under study are induced periodically by external means, e.g., an electric field, temperature gradient or mechanical strain.\ As the optical linear response of the medium is governed by its dielectric constant, changes in the latter induce changes in its reflectance and transmittance, which can be detected using phase-sensitive amplifiers locked to the frequency of the external perturbation.\ PR is a particular form of contactless electro-reflectance (ER) \cite{Aspn80, Sera65}, in which built-in electric fields, associated to space-charge regions at junctions and surfaces of the sample, are perturbed periodically by means of a chopped pump-light beam.\ Signatures in modulation spectroscopy appear at energies corresponding to critical points in the electronic structure of the sample.\ Modulation spectroscopy, contrarily to the case of luminescence, is considered a form of absorption-spectroscopy, as it can probe unoccupied electronic states of the samples under test.\ Since its proposal \cite{Sera65} modulation spectroscopy has been extensively used in the characterization of semiconductor materials and devices, including bare material \cite{Walu00}, homo- and heterojunctions \cite{Yin90}, and different sorts of nanostructures \cite{Misi12}.\ The application of PR to monolithically integrated optoelectronic devices, like multijunction solar cells, has been demonstrated before \cite{Fuer16, Cano10}.\ The main difficulty in the characterization of such two-terminal devices is the independent optoelectronic diagnosis of each of the series-connected subcells forming the device.\ The development of spectroscopic techniques capable of complementing conventionally used ones, like quantum efficiency (QE) or photoluminescence (PL), is therefore of great interest.\ With this aim, we have explored the possibility of extending the capabilities of PR when applied to multijunction solar cells (MJSC) under different bias conditions.\ In particular, we will show how to activate selectively the response of individual subelements of a monolithically integrated device when acting on a different subelement with the pump-light beam.\ In this manner it is possible to circumvent limitations on the information depth attainable from conventional PR, related to the optical absorption depth in the material of the modulating pump beam and the diffusion length of thereby photogenerated minority carriers.\ In previous studies luminescent coupling between subelements in a monolithic device was identified as an effective mechanism for transmitting the modulating perturbation to otherwise inaccessible parts of the sample \cite{Fuer16}.\ Here we propose yet a different method for the transmission of the modulation based on the electric coupling between subelements, thus not restricted by the internal quantum yield of radiative recombination process in the pump-beam-absorbing material.

\begin{figure}
\includegraphics{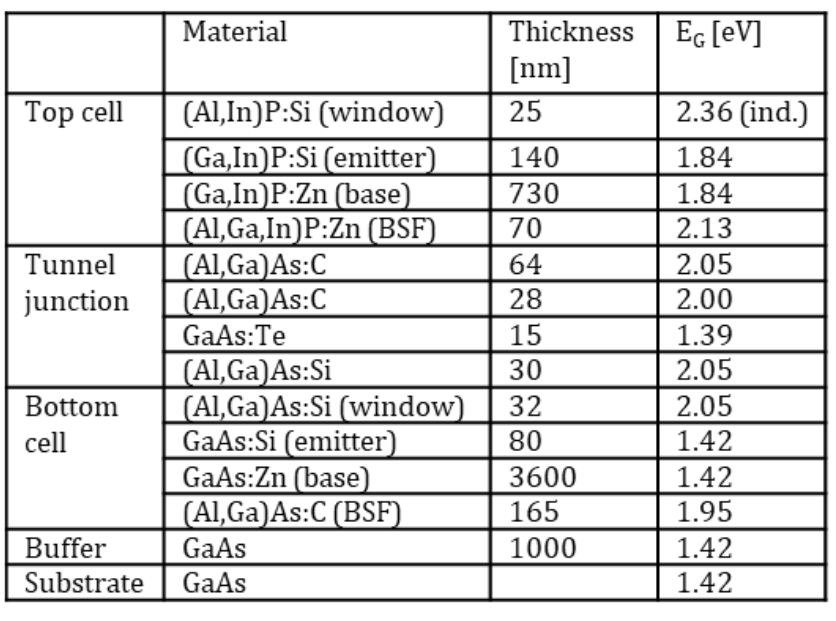}
\caption{\label{fig1:wide} Layer structure of the MOVPE-grown GaInP/GaAs double-junction solar cell structure under study, with layer role (\textit{BSF} stands for back-surface-field layer), nominal thickness and fundamental bandgap values (\textit{ind.} reads indirect)}
\end{figure}

A piece of lattice matched, metalorganic vapor-phase epitaxy (MOVPE)-grown GaInP/GaAs double-junction solar cell has been used in this study, with the layer structure indicated in Fig.~\ref{fig1:wide}.\ The sample did have neither AR-coating nor front metal grid.\ It was provided instead with a Ni-pad sideways of the front active area and mounted onto a Cu-disk with In in order to enable electric contacts.\ The external QE of the device revealed a good current matching between top- (TC) and bottom cells (BC) at around $10$ mA/cm{$^2$}.\ The estimated conversion efficiency of the device is about 25\% under one-sun AM1.5D illumination.\ Room-temperature PR measurements were performed using the 325 nm line of a HeCd laser (Oriel, 15 mW) and the 814 nm line of a of variable intensity solid-state laser ($\mu$LS) as pump beams, chopped at 777 Hz.\ For the probe beam, the light of a 250 W QTH-lamp was dispersed with a 1/8 monochromator (Oriel).\ Direct reflectance of the probe beam, containing average (dc) and modulated (ac) components, were detected with a Si-photodiode connected to a preamp (Keithley) and a lock-in amplifier (Stanford).\ The PR signal is defined as the spectral ac/dc ratio.\ Supporting numerical calculations were run on TCAD device simulator Atlas (Silvaco), using finite elements to self-consistently solve drift-diffusion and Poisson differential equations, non-local models for tunnel junctions and generalized matrix methods for photogeneration, as described before \cite{Ocho16}.\ Fermi-Dirac statistics were used and both radiative and non–radiative recombination processes considered.

\begin{figure*}
\includegraphics{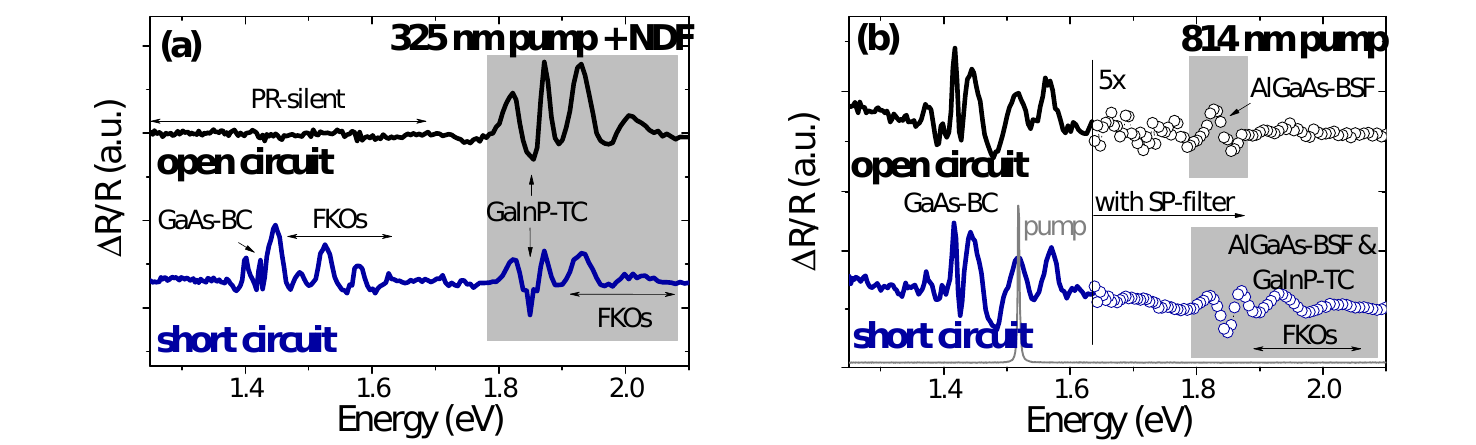}
\caption{\label{fig2:wide} (a) Comparison of PR spectra obtained in open- (top) and short-circuit conditions (bottom) under 325 nm pump using a neutral density filter, such that modulation of BC due to luminescence coupling lies below the background level in open-circuit. In short circuit electric coupling enables ER-modulation in the BC. (b) The same comparison under 814 nm pump. Open symbols identify the spectral range measured with a SP-filter at the detector; the thin solid line represents the spectral distribution of the 814 nm source, in arbitrary units. Spectra are offset in both plots for clarity.}
\end{figure*}

Previous PR experiments on identical samples with 325 nm (3.8 eV) pump beam had revealed signatures from both the GaInP-TC and the GaAs-BC \cite{Fuer16}.\ The nature of the modulation mechanism of the BC and the appearance of related signatures in PR spectra is not straightforward.\ Considering layer thicknesses and absorption coefficients, full absorption of the 325 nm pump beam is expected at the TC (870 nm, $\alpha_{GaInP}=7.9\times10^5 cm^{-1}$) and its AlInP-window layer (25 nm, $\alpha_{AlInP}=6.7\times10^5 cm^{-1}$) under normal incidence, according to numerical calculations.\ No modulation and, consequently, no related PR-signatures were thus expected from the BC, provided no significant shunt bypasses the tunnel junction connecting both subcells.\ We have previously identified luminescent coupling (LC) \cite{Stei13} as the modulation mechanism responsible of PR signatures of low gap sub-cells under TC-excitation and proposed a method to quantify the LC-impact in form of a measurable photovoltage \cite{Fuer16}.\ Under LC, radiative recombination events occurring at the TC, following photo-excitation with the pump light, generate subsidiary photons that can be re-absorbed at the BC, thereby generating additional photocurrent and effectively transmitting the pump-induced photoexcitation into deeper layers within the structure, beyond the penetration depth of the pump beam itself.\ In this manner, the effect of the modulation is not necessarily limited by the diffusion length of minority carriers created in the volume of pump light absorption \cite{Moch95, Moty07}.\ Furthermore, LC permits the transmission of the modulation over potential barriers for diffusive transport, like tunnel junctions in the case of MJSC in open-circuit conditions.\ Our results revealed a direct correlation between the observed LC-induced photovoltage recorded during PR measurements, and the onset of PR signatures from the BC.\ BC-related PR-signatures are thus unambiguously associated to LC-induced PR, with the TC acting as subsidiary pump excitation source.\ The magnitude of LC can be controlled by means of a neutral density filter (NDF) attenuating the pump beam intensity exciting the TC.\ In this manner, we can selectively activate/deactivate the appearance of BC-related PR-signatures above/below the background noise level.\ The top spectrum of Fig.~\ref{fig2:wide}a shows the recorded PR signal obtained in open circuit conditions using a NDF at sufficiently low pump-intensity as to keep the BC PR-silent, while signatures related to the GaInP-TC bandgap at 1.85 eV and related Franz-Keldysh oscillations (FKOs) are readily visible.\ We have performed additional PR measurements using the 814 nm (1.52 eV) pump beam.\ As the pump-photon energy is now lower than the bandgaps of all layers forming the TC, the 814 nm pump beam is largely transmitted through and absorbed at the BC (absorption at the thin tunnel junction is neglected).\ Fig.~\ref{fig2:wide}b (top panel) shows the spectrum obtained in open circuit conditions with 814 nm pump, divided in two ranges: the high energy range (dotted) was obtained placing a short-pass (SP) filter in front of the detector, cutting off photons of wavelengths beyond 750 nm and thereby impeding the entrance of laser scattering in the detector; the low energy range (solid line) was obtained without pass filter.\ It is worth recalling that the pump-photon energy does not in principle set any fundamental upper limit for detection in PR, provided the pump is capable of generating photovoltage in the sample as result of interband transitions \cite{Fuer17}: the pump-induced photovoltage affects the entire electronic structure as result of band bending, including unoccupied states not directly reached by pump photons that can still be probed at the corresponding energy above $E_{pump}$.\ It can be observed that the top PR-spectrum in Fig.~\ref{fig2:wide}b contains signatures showing up at energies higher than that of the pump beam (whose spectral distribution is shown in the graph for reference): the fading tail of Franz-Keldysh oscillations (FKOs) associated to the 1.42 eV critical point of the GaAs-BC and a signature at 1.82 eV (some 30 meV below the gap GaInP-TC obtained with 325 nm pump), which was buried in the background of filter-free measurements.\ The origin of the 1.82 eV signature is attributed to the modulation of the interface between the BC GaAs-base and its AlGaAs-BSF, as discussed later.\ The lower panels in Figs.~\ref{fig2:wide}a and b show PR spectra obtained in short-circuit conditions under 325 and 814 nm pump illumination, respectively.\ Comparing short- and open-circuit cases, we can observe that signatures from the subcell at which the pump-beam-induced photovoltage is generated (the TC for 325 nm pump and the BC for 814 nm pump) remain largely unaltered.\ On the other hand, significant changes are observed in the energy ranges of the complementary subcell, as result of the electric coupling in short-circuit conditions.\ In the case of 814 nm pump, the photovoltage generated by pump illumination at the BC ($\approx440$ mV according to photovoltage measurements not shown) must be cancelled out in short circuit by reverse biasing the TC in the same magnitude right at the modulating frequency of the pump laser.\ As a consequence, the TC is electro-modulated upon the action of photo-modulation of the BC. The effect is revealed by the presence of the GaInP-TC signature at 1.85 eV and the associated FKOs, which appear superimposed to the BC-BSF signature at 1.82 eV observed in open-circuit.\ The same occurs when illuminating the device in short circuit with 325 nm pump: the TC develops $\approx145$ mV forward bias as a result of photogeneration when the laser is attenuated with NDF (1160 mV without NDF), reverse biasing the BC accordingly at the same frequency.\ The probe beam thus records the effects of both modulation mechanisms acting on different subcells, in the form of a composed electro-photoreflectance spectrum.\ Furthermore, the results of Fig.~\ref{fig2:wide} demonstrate that the mechanism of electroreflectance enabled by electric coupling operates reversibly between top and bottom subcells when selectively exciting one of them with the corresponding pump beam.\ Such combined modulation mechanism thus permits to probe different subcells independently in any kind of monolithic multijunction device.\ The relevance of the proposed method in the field of optoelectronic characterization of monolithically integrated devices can be inferred from the above results.\ Electric coupling of subelements under short-circuit conditions permits the activation of the response from individual subelements not affected by the perturbing agent and therefore expected to be PR-silent. Electric coupling thus enables the transmission of the modulating perturbation to other subelements beyond the limits imposed by (i) the absorption depth of the pump light, (ii) the quantum yield of radiative recombination eventually activating luminescent coupling, and (iii) the diffusion length of minority carriers photogenerated by the pump beam.\ It should be noticed that the application of ER to the complete monolithic device by means of external bias modulation as an alternative method does not permit the individual response activation of silent subelements as illustrated in Fig.~\ref{fig2:wide}, as the modulating periodic voltage bias applied to the contacts will affect all subelements and space-charge regions present in the device. 

\begin{figure}
\includegraphics{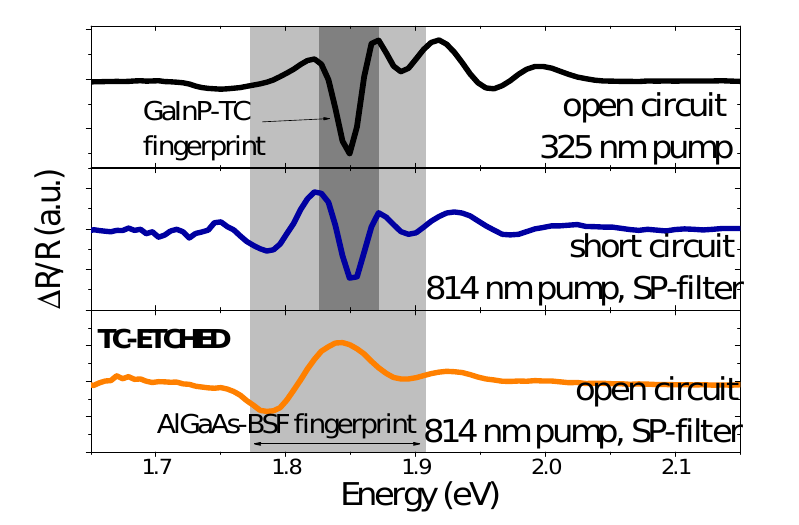}
\caption{\label{fig3:wide} PR spectra in the energy range of the TC under 325 nm in open circuit (top); under 814 nm pump in short circuit (middle); and under 814 nm pump in open circuit, after chemical removal of TC and tunnel junction (bottom).\ Shaded areas highlight signatures of the GaInP-TC at 1.85 eV (dark) and the AlGaAs-BSF at 1.82 eV (light).}
\end{figure}

We turn now back to the origin of the 1.82 eV signature.\ Fig.~\ref{fig3:wide} shows PR-measurements carried out on a twin sample over the energy range of the TC with 325 nm pump in open circuit conditions (top panel), together with two additional spectra recorded under 814 nm laser pump: with the device in short circuit before chemically etching the TC and the tunnel junction by HCl dipping (medium panel); and in open circuit after the TC and tunnel junction had been etched away (bottom panel).\ No NDF-filter was used for the spectrum obtained with 325 nm pump in order to permit BC modulation by LC and the eventual appearance of BC-related signatures in the probed range.\ As it can be observed, measurements recorded under 325 nm pump in open circuit (top) and under 814 nm pump in short-circuit (medium) are nearly identical, indicating that the same critical points and related FKOs are probed on a similar basis.\ The spectrum is clearly modified after etching the TC away, as shown in the bottom panel:\ the narrow signature peaking at 1.85 eV is eliminated and both the intensity and period of associated FKOs are modified.\ As TC removal does not result in a featureless spectrum, it is therefore concluded that the PR spectra of complete, non-etched samples, contain signatures of at least two overlapping contributions in the same energy range, namely: (i)	a narrow signature at 1.85 eV, unambiguously attributed to the TC, with enhanced FKOs; and (ii) a broader signature at 1.82 eV, together with FKOs of minor amplitude, whose origin must be attributed to the BC.

\begin{figure*}
\includegraphics{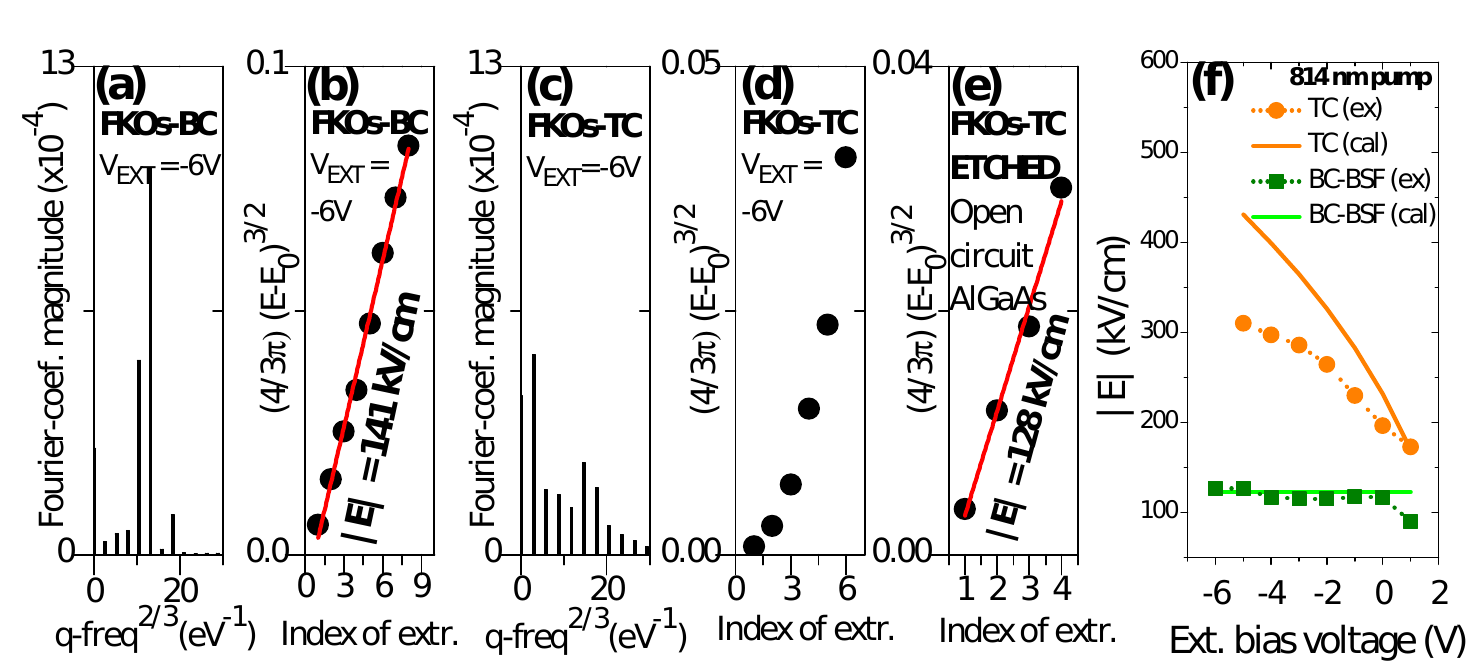}
\caption{\label{fig4:wide} FKO-analysis of BC and TC energy ranges of PR spectra obtained under 325 nm pump under different applied voltage bias; (a) \& (c) Fast-Fourier Transform analysis of FKOs of BC and TC, respectively, representing coefficient magnitudes as a function of quasi-frequencies; in (a) a single dominant frequency is observed, attributed to BC pn-junction, whereas in (c) distributed frequencies contribute in the TC and BC-BSF energy range. (b) \& (d):\ plots of $\left(E-E_{0}\right)^{3/2}$ vs index of oscillation extrema of FKOs of BC and TC energy ranges, respectively; (b) includes a linear regression and maximum electric field estimation of 141 kV/cm for the BC-junction; (d) reveals a non-linear dependency resulting from concurrent fields. (e) $\left(E-E_{0}\right)^{3/2}$ vs index of oscillation extrema and linear regression obtained after TC and tunnel junction removal. The estimated field intensity at GaAs/AlGaAs is $128$ kV/cm. (a-d) measured under -6 V external reverse bias; (e) measured in open circuit. (f) Experimental and calculated field intensity as a function of external voltage bias at TC-pn- and BC-BSF-junctions under 814 nm pump in short-circuit.}
\end{figure*}

The interpretation of these results is further supported by the analysis of FKOs of etched and non-etched samples under different illumination and externally applied voltage bias, over the ranges of the respective subcells.\ The FKOs period is a direct measure of the built-in field intensity in the probed space-charge region \cite{Shen95, Aspn73, Shen90, Glos87, Bott88, Wang95}.\ We recall that FKOs in PR spectra at energies above a critical point can be described by \cite{Shen95}: 

\begin{equation}
	\frac{\Delta R}{R} \propto cos \left\{ \frac{4}{3} \frac{\left(E-E_{0}\right)^{3/2}}{\hbar \theta} +\varphi \right\}, 
\end{equation}

where ~$(\hbar\theta)^3 = e^2\hbar ^2 F^2/(2\mu)$ is the electro-optic energy, $\varphi$ is a phase factor related with the dimensionality of the critical point $E_0$, $F$ is the intensity of the electric field, and $\mu$ is the reduced interband effective mass in the direction of the field.\ Two forms of FKO period analysis are usually employed:\ the analytical method relies on the fact that the oscillation extrema of FKOs stemming from a single field source satisfy:

\begin{equation}
	n\pi = \frac{4}{3} \frac{\left(E_n-E_{0}\right)^{3/2}}{\hbar \theta} +\varphi , 
\end{equation}

where $n$ is the integer index of the extrema and $E_n$ the energy of the $n$th extremum, and the extrema are therefore directly proportional to $(E-E_{0})^{3/2}$ with a slope inversely proportional to the magnitude of the field \cite{Shen95}.\ The second method is based on the Fourier analysis of the oscillating spectrum and allows the identification of multiple contributions to the FKOs in the form of distributed quasi-frequencies \cite{Wang95}.\ Fig.~\ref{fig4:wide} shows results of both types of analysis, as recorded from a complete sample under 325 nm pump and -6 V reverse bias.\ Figs.~\ref{fig4:wide}a and b refer to FKOs recorded in the range of the BC and show the distribution of Fourier coefficients as a function of quasi-frequency and the linear regression obtained from the analytical method, respectively.\ As it can be observed, a single frequency contribution dominates the Fourier spectrum, in accordance with a single source of electric field located at the BC-junction.\ On the other hand, Figs.~\ref{fig4:wide}c and d, corresponding to the energy range of the TC, reveal a distribution of frequencies in the Fourier spectrum and a non-linear dependency of $(E-E_{0})^{3/2}$ vs. index relation, unambiguously indicating concurrent field sources, attributed to GaInP-TC and AlGaAs-BSF.\ The linearity predicted by the analytical method is recovered from the remaining FKOs after removing the TC and tunnel junction, as shown in Fig.~\ref{fig4:wide}e.\ The origin of the single contribution to the field is consequently attributed to the AlGaAs-BSF at the rear side of the BC, as pointed out previously.\ The calculated field intensity at the GaAs/AlGaAs junction in open-circuit conditions is 128 kV/cm, in excellent agreement with numerical calculations.\ The individual subcell study is completed with the calculation of the electric field intensities present at the TC pn-junction and BC-BSF interface as a function of external voltage bias.\ From the distinct evolution under external reverse bias of the two main contributions to the quasi-frequency spectrum of Fourier-transformed FKOs, both fields can be quantified.\ Results obtained from PR spectra under 814 nm pump in short-circuit are shown in Fig.~\ref{fig4:wide}f.\ We emphasize that both contributions are simultaneously accessible over the energy range probed thanks to the electric coupling between subcells, as otherwise the TC would remain PR-silent.\ The expected square-root dependency of field intensity with voltage of the TC-junction and the concurrent invariance of the weak BC-BSF junction are readily visible.\ Solid lines represent numerical calculations of the corresponding field intensities at given voltages.\ Deviations from theoretical values can be accounted for as (i) the applied bias does not drop exclusively at the TC space-charge region but is rather distributed over contacts and junctions of the entire device, and (ii) other type of non-idealities, like resistive losses associated to current flow in short-circuit or charging/discharging of interface defects at heterojunctions have not been considered.\ Nevertheless, the method hereby proposed permits to interrogate on such nonidealities in monolithically integrated optoelectronic devices.\ Such studies are underway.\ It should be pointed out that the nature of the modulation at the BSF and the TC in complete devices is different and depends on the wavelength of the pump and the electric coupling of the device.\ For the case of 325 nm pump in open circuit, the TC is directly modulated by photogenerated carriers, whereas all BC-related features, included its BSF, are activated via LC.\ For 814 nm pump illumination, TC related signatures show up with the device in short circuit in the form of an ER spectrum, whereas BSF modulation with sub-bandgap pump light is enabled across the heterojuntion interface.\ This type of heterojunction-mediated sub-bandgap modulation has been reported before, \textit{e.g.}, by Lee \textit{et al.}\ \cite{Lee88}.\ Finally, it is worth commenting that the attribution of the PR-signature at 1.82 eV  to the AlGaAs-BSF does imply a deviation in the actual composition of the BSF layer, from the nominal 40\% Al-content (atomic, corresponding to a bandgap of 1.95 eV) to a lower Al-content of 31.3\% (1.82 eV bandgap), thereby underlying the potential of PR as a diagnostics tool in the non-destructive characterization of buried layers.

In summary, it has been shown that electric coupling between subcells of monolithically grown devices in short circuit conditions allows their simultaneous and independent characterization in the form of combined photo- and electroreflectance spectra.\ Alternatively, LC enables full device modulation in the form of pure photoreflectance spectra.\ The combination of such forms of modulation spectroscopy may be of interest for the characterization of composed optoelectronic devices and particularly in the field of MJSC diagnosis.   

Financial support from Comunidad de Madrid (S2013-MAE-2780), the Ministry of Economy and Competitivity (TEC2015-64189-C3-1-R), the European Commission (COST-action MP1406, DFM), and EU-Horizon 2020 research and innovation program (Marie Sklodowska-Curie grant agreement No 656208, EB) is acknowledged.

\bibliography{PR-DJSCNEWv1}

\end{document}